\documentclass[preprint,brazil,english,sort&compress]{elsarticle}
\pdfoutput=1
\usepackage[T1]{fontenc}
\usepackage[utf8]{inputenc}
\usepackage{babel}
\usepackage{units}
\usepackage{amsmath}
\usepackage{amssymb}
\usepackage{graphicx}
\usepackage[unicode=true,
 bookmarks=true,bookmarksnumbered=false,bookmarksopen=false,
 breaklinks=false,pdfborder={0 0 1},backref=false,colorlinks=false]
 {hyperref}
\hypersetup{pdftitle={Individual discrimination of freely swimming pulse-type electric fish from electrode array recordings},
 pdfauthor={Paulo Matias, Jan Frans Willem Slaets, Reynaldo Daniel Pinto},
 pdfsubject={Signal processing and machine learning supporting neuroethology research},
 pdfkeywords={Neuroethology, Electric organ discharge, Classification, Dual-tree complex wavelet packet, Support vector machine, Continuity constraint}}

\makeatletter

\providecommand{\tabularnewline}{\\}

\journal{Neurocomputing}

\biboptions{compress}

\DeclareMathOperator*{\argmin}{\arg\!\min}

\makeatother

\begin{document}

\title{Individual discrimination of freely swimming pulse-type electric
fish from electrode array recordings}

\author[ifsc]{Paulo Matias\corref{corauthor}}

\ead{paulo.matias@usp.br}

\author[ifsc]{Jan Frans Willem Slaets}

\author[ifsc]{Reynaldo Daniel Pinto}

\cortext[corauthor]{Corresponding author. Tel.: +55 16 33738090; fax: +55 16 33739879.}

\address[ifsc]{Department of Physics and Interdisciplinary Science, \foreignlanguage{brazil}{São Carlos}
Institute of Physics, University of \foreignlanguage{brazil}{São Paulo}, \foreignlanguage{brazil}{Av.
Trabalhador Sancarlense} 400, \\13560-970 \foreignlanguage{brazil}{São
Carlos}, SP, Brazil}
\begin{abstract}
Pulse-type weakly electric fishes communicate through electrical discharges
with a stereotyped waveform, varying solely the interval between pulses
according to the information being transmitted. This simple codification
mechanism is similar to the one found in various known neuronal circuits,
which renders these animals as good models for the study of natural
communication systems, allowing experiments involving behavioral and
neuroethological aspects. Performing analysis of data collected from
more than one freely swimming fish is a challenge since the detected
electric organ discharge (EOD) patterns are dependent on each animal's
position and orientation relative to the electrodes. However, since
each fish emits a characteristic EOD waveform, computational tools
can be employed to match each EOD to the respective fish. In this
paper we describe a computational method able to recognize fish EODs
from dyads using normalized feature vectors obtained by applying Fourier
and dual-tree complex wavelet packet transforms. We employ support
vector machines as classifiers, and a continuity constraint algorithm
allows us to solve issues caused by overlapping EODs and signal saturation.
Extensive validation procedures with \textit{Gymnotus} sp. showed
that EODs can be assigned correctly to each fish with only two errors
per million discharges.\end{abstract}
\maketitle

\section{Introduction}

\hyphenation{Gymnotus}

Pulse-type weakly electric fishes such as \textit{Gymnotus} sp. are
known for emitting electric organ discharges (EODs) used for electrolocation
and electrocommunication purposes \citep{pmid13654750,Caputi2002493}.
Because of the stereotyped nature of the electrical waveforms produced
by these fish, research on electrocommunication typically focus on
analyzing measurements derived only from the occurrence instant (timestamp)
of EODs, for example its discrete difference --- the inter-pulse interval
(IPI) \citep{Westby.lat.comm,ISI:000248304900011,JFB:JFB3448}. Few
organisms allow the non-invasive examination of electrophysiological
signals produced by a complex internal neuronal network as offered
by this simple communication mechanism based on trains of electrical
pulses.

Recording from freely swimming fish is challenging and traditional
techniques based in arrays of electrodes fixed in the aquarium are
still employed \citep{Westby1975192,jun2012precision,10.1371/journal.pone.0084885}.
The idea is similar to using an electrode array to record from neurons
in the central nervous system of an animal \citep{Gross1977101,Kipke1214707},
however, here the spiking ``neurons'' are not stationary in space
and the problem resembles, but it is even more complicated than, that
of recording from arrays of electrodes which present drift along time
\citep{Snider1998155}. Techniques allowing the precise detection
and discrimination of EODs emitted by dyads or groups of freely moving
fish with a minimum disturbance are invaluable tools for neuroethological
research \citep{ETH:ETH2022,perrone2009social,Arnegard07072005,JFB:JFB3448,ISI:000248304900011},
because they allow to study a plethora of social communication circumstances
in a naturalistic setup. However, currently there is a lack of computational
tools capable of accurately identifying the individual that emitted
each pulse recorded during those experiments.

In principle, employing machine learning techniques to address the
individual discrimination problem would be feasible since EOD waveforms
vary from one fish to another \citep{McGregor1992977}. But although
the distinct waveforms of different individuals have been the object
of study in reports on EOD variations related to geographical origin
\citep{gallant2011signal}, characteristics of the waveforms have
seldom been exploited for recognizing fish. There are also reports
on changes of the waveform due to developmental transitions in juvenile
fish \citep{crampton2005nesting}, but the EOD of a certain individual
does not change by factors other than fish movement \citep{hopkins1986temporal,jun2012precision}
during our experiment's time frame (a few hours).

Most of the existing works try to discriminate individuals by employing
non-automatic procedures involving visual inspection of EOD pulse
amplitude and duration \citep{Arnegard07072005}, sometimes aided
by video recordings that allow inferring fish position, which is then
manually correlated to pulse amplitude and polarity changes \citep{JFB:JFB3448}.
Still, those methods are time-consuming and often applied solely to
a few minutes of experimental data, and therefore they do not produce
sufficient input to enable statistical and information-theoretic approaches
\citep{10.1371/journal.pone.0084885} to study spike train coding
\citep{PhysRevLett.80.197}.

An automated method was proposed \citep{ISI:000248304900011} that
first stored two template EODs, each one from a single fish, then
computed the cross-correlation between every EOD acquired in a dyad
experiment and each of the templates. Nevertheless, that approach
had difficulties when EODs had similar pulse width or when both fish
fired pulses almost at the same time.

In this paper we introduce a computational method able to discriminate
fish within dyads surpassing the just mentioned limitations, given
as input measurements from electrodes placed at fixed positions in
an aquarium tank. Our method is essentially a two pass algorithm.
First, signal portions strongly believed to contain an EOD emitted
by a single fish are classified by a support vector machine \citep{cortes1995support},
based on a normalized feature vector obtained by applying Fourier
and dual-tree complex wavelet packet \citep{Bayram:2008:DCW:2198040.2204301,Weickert:2009:AWP:1653617.1653625}
transforms. Then, the fact that waveforms vary continuously during
fish movement is exploited to find EODs inside signal segments which
might contain discharges from both fish.

We also perform an extensive validation procedure during which dipoles
are attached directly to each fish to capture the EODs whose timestamps
are then compared with the results of the developed algorithm in order
to evaluate its error rate, which we estimate as being only two parts
per million.

The paper is organized as follows. Section \ref{sec:Experimental-methods}
introduces the experimental methods, explaining how measurements are
carried, and also detailing the procedure we perform to validate our
method and evaluate its accuracy. Section \ref{sec:Discrimination-methods}
describes both the discrimination method already present in literature
and our proposed algorithm. Finally, results are presented in Section
\ref{sec:Results} and conclusions are drawn in Section \ref{sec:Conclusions}.

\section{Experimental methods\label{sec:Experimental-methods}}

\begin{figure*}
\begin{centering}
\includegraphics[width=1\textwidth]{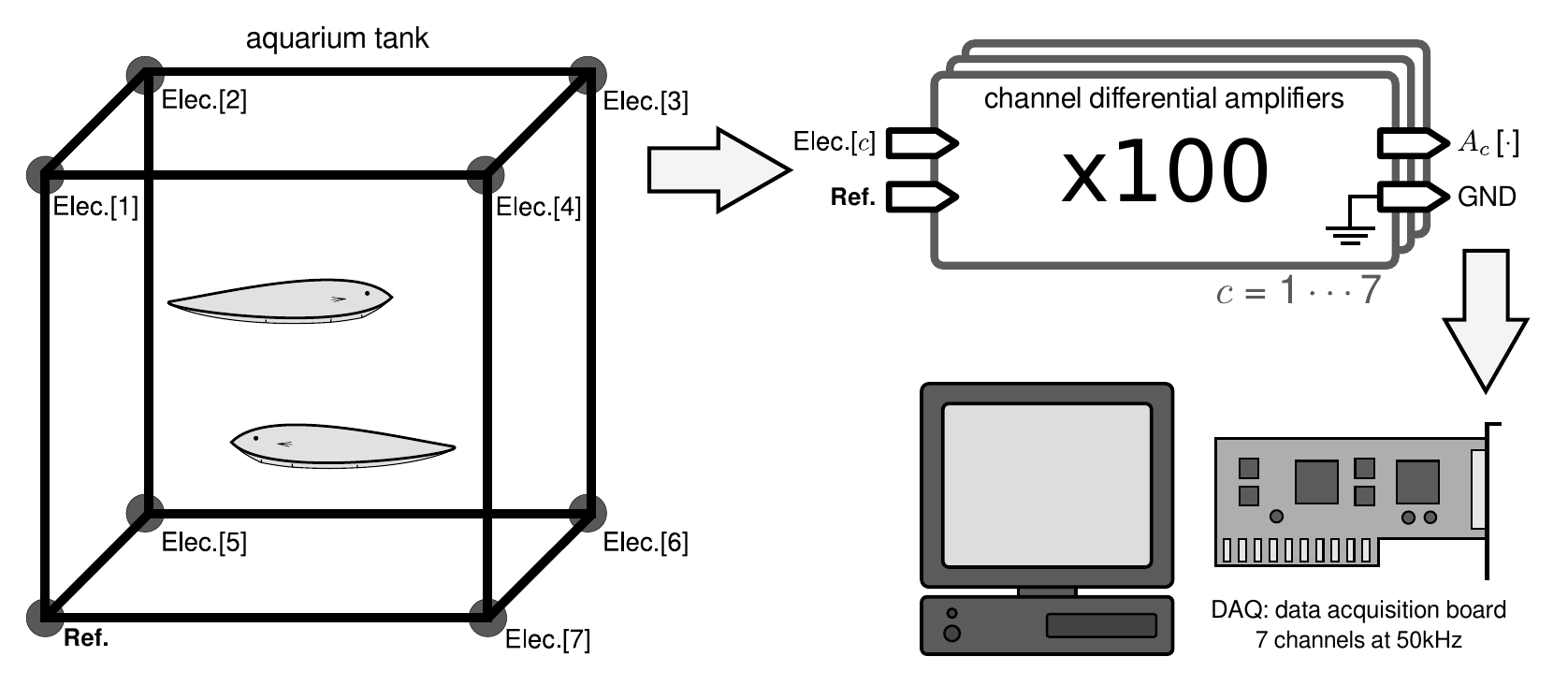}
\par\end{centering}

\caption{\label{fig:AquariumBlockDiagram}A pair of fish is placed in an aquarium
tank containing eight electrodes in contact with water. One of the
electrodes is chosen as a reference electrode (ref.), in respect to
which the voltage of all other electrodes is differentially amplified
with a 100 times gain. Signals sampled at 50 kHz are then collected
by a computer using a data acquisition board.}
\end{figure*}

\subsection{Experimental setup}

Our experimental setup is illustrated in Figure \ref{fig:AquariumBlockDiagram}.
Measurements are taken in a 64 liters glass aquarium tank containing
eight stainless steel electrodes, located at the vertices of a 40
cm sided cube. The aquarium is mounted inside a Faraday cage to reduce
the induction of external electrical noise. Electrodes with a diameter
of 0.2 mm are inserted through the silicon glue at the corners of
the aquarium, having about 1 to 2 mm of length in contact with the
water. One of the electrodes is chosen as a reference, in respect
to which the voltage of the other seven electrodes is differentially
amplified 100 times using LM308 operational amplifiers. Once amplified,
the seven signals are digitized at a sampling rate of 50 kHz with
a resolution of 12 bits by a National Instruments PCI MIO-16-E1 data
acquisition board and stored in a personal computer.

When fish gets too close to the reference electrode, an exceeding
voltage may be produced between the reference and other electrodes,
saturating signals recorded from all of them at the same time. To
avoid this issue, we fixed a piece of nylon tulle to the aquarium
glass near the reference electrode, preventing fish from reaching
it.

Our aquarium geometry and electrode placement is the same adopted
in \citep{10.1371/journal.pone.0084885}. It is easy to replicate
and provides an adequate signal to noise ratio (SNR) in at least one
electrode independently of the fish position. However, nothing precludes
the algorithm described in this paper from being used with other geometries,
such as the round aquarium with multiple reference electrodes as described
in \citep{jun2012precision}.

\subsection{Experimental procedure}

Experiments with a fish dyad are comprised by two steps which can
be carried in any desired order. We call one of these steps the training
stage, which consists of placing apart in the aquarium each individual,
in turn, for some minutes, during which the EODs of the freely swimming
fish are collected. Typically, 15 minutes of acquisition are enough
to collect on the order of $10^{5}$ EODs, sufficient for training
and testing our classifier. In order to acquire good quality labeled
training and testing data, covering the system dynamics over most
of the operating range, it is important to get the fish to swim around
all aquarium. In \textit{Gymnotus} sp., this usually occurs naturally,
as the fish has a tendency to explore the surroundings when it is
moved to a different environment \citep{10.1371/journal.pone.0084885}.
The experimenter can also arouse an inactive animal by mechanically
disturbing the aquarium.

The other step, which we call the main experiment, consists of placing
both fish at the same time inside the aquarium. Data acquired in this
step can be discriminated by our algorithm, outputting a list of the
occurrence instants of EODs emitted by each fish. These instants are
the final product of our method, and can be analyzed and studied in
order to research new behavior and codification schemes occurring
in fish electrocommunication.

\subsection{Validation procedure\label{sub:Validation-procedure}}

To evaluate if our discrimination algorithm gave accurate results,
we conducted experiments where electrodes were attached next to each
fish and recorded in addition to the fixed electrode array already
present in the aquarium. To keep the electrodes near the fish, the
wires of the electrodes were intertwined to a nylon tulle, which was
wrapped around the animals. Therefore, fish movement was fairly restrained
during these tests, demanding that we manually moved the individuals
around the aquarium, by pushing and pulling the wires, to simulate
the position displacements of a freely swimming fish.

Instants of EOD emission could thus be directly obtained by applying
a simple threshold to the signal captured from the electrode tied
to each individual. These direct measurements were then compared with
the output of our discrimination algorithm to which only the fixed
electrode measurements were supplied as inputs.

Six different \textit{Gymnotus} sp. dyads were used for carrying this
validation procedure. For each dyad, two experiments were made. In
the first one, both fish were restrained by nylon tulle cover and
manually moved around the aquarium. Then, during the second experiment,
one arbitrarily chosen individual of the dyad was freed and allowed
to swim, while the other fish was kept enclosed near its electrode,
and hence had its EOD instants directly measured.

\section{Discrimination methods\label{sec:Discrimination-methods}}

\subsection{Cross-correlation method\label{sub:Cross-corr-method}}

Before discussing our algorithm, we briefly introduce our implementation
of the cross-correlation method already described in literature \citep{ISI:000248304900011},
which we will use for comparison purposes.

Given a signal $A\left[\cdot\right]$ containing an unlabeled EOD,
the method computes the cross-correlation $T_{k}\star A$ for both
$k=1$ and $k=2$. Each value of $k$ corresponds to one fish of the
dyad, and $T_{k}$ is the signal template of such fish.

\[
\left(T_{k}\star A\right)\left[i\right]=\sum_{j=0}^{l_{w}-1}T_{k}\left[j\right]\cdot A\left[i+j\right]
\]

Where $l_{w}$ is the maximum length of an EOD in number of samples.
The method then computes $\mathrm{MaxCorr}_{k}$, the maximum absolute
value of the cross-correlation for the $k$-th fish, considering all
the possible EOD starting instants $i$ inside the signal $A\left[\cdot\right]$.

\[
\mathrm{MaxCorr}_{k}=\max_{i\in[-l_{w},l_{w}]}\;\left|\left(T_{k}\star A\right)\left[i\right]\right|
\]

Then the $A\left[\cdot\right]$ signal is classified as containing
an EOD from ``fish 1'' if $(\mathrm{MaxCorr}_{1}-\mathrm{MaxCorr}_{2})\ge\mbox{\ensuremath{\mathrm{DecThreshold}}}$.
Otherwise, it is classified as ``fish 2''. The decision threshold
($\mbox{\ensuremath{\mathrm{DecThreshold}}}$) did not exist in the
original method. We introduced it as a parameter that can be varied
to plot receiver operating characteristic (ROC) curves (e.g. Figure
\ref{fig:Results_ROC}). Setting the threshold to zero recovers the
results of the original method.

We choose templates $T_{k}$ by exhaustive search over all possible
pairs of EODs extracted from subsets of $10^{3}$ labeled pulses selected
at random from those collected during the training stage. We normalize
them such that $\max_{j}\left|T_{k}\left[j\right]\right|=1$. We select
the templates which provide the best classification accuracy on the
other pulses of the training data, i.e. those not contained in the
subsets where the template search is done.

An issue of the cross-correlation method is the low accuracy it presents
when both fish of a dyad emit EODs of almost the same pulse duration.
Also, the method does not include any means of dealing with overlapped
EODs. Thus, our main challenges when implementing a new algorithm
were to address these two points --- classifying signals according
to their fine structure even though they vary with fish position,
and discriminating EODs even when both fish fire almost at the same
time.

\subsection{Proposed algorithm\label{sub:Proposed-algorithm}}

Our discrimination algorithm works on signal segments which may contain
either one or two EODs, respectively from a single fish or from both
fish of a dyad. Subsection \ref{sub:EOD-segmentation} describes how
these segments are obtained from the acquired signals. Then, the algorithm
conceptually consists of two main steps: the classification of single
fish segments and the dissociation of signal segments containing EODs
from both fish. The core of the first step is a support vector machine
(SVM) classifier. The extraction and the selection of features which feed
the SVM are detailed, respectively, in Subsections
\ref{sub:Feature-extraction} and \ref{sub:Feature-selection}, and
the application of the SVM to classify segments containing a single
EOD is depicted in Subsection \ref{sub:Supervised-classifier}. The
second step consists of applying a continuity constraint to allow
the discrimination of two EODs present in the same segment, which
is portrayed in Subsection \ref{sub:Continuity-constraint}.

\subsubsection{EOD segmentation\label{sub:EOD-segmentation}}

Data both from the training stage and from the main experiment are
segmented before analysis. The purpose of this procedure is to detect
EODs, distinguishing them from background noise, and to estimate the
time span of EOD activity on each detection.

Signals $A_{c}\left[\cdot\right]$ from all channels $c$ are passed
by a low-pass finite impulse response filter (Hamming window, 11-tap,
$2500$ Hz cutoff), differentiated and squared. If the sum $I\left[\cdot\right]$
of the resulting signals surpasses a threshold value $t_{d}$, a segment
of electric organ activity is detected.

\[
F_{c}\left[\cdot\right]=A_{c}\left[\cdot\right]\ast\text{Filter}\left[\cdot\right]
\]

\[
\Delta F_{c}\,\left[i\right]=F_{c}\left[i+1\right]-F_{c}\left[i\right]
\]

\[
I\left[i\right]=\sum_{c}\left(\Delta F_{c}\,\left[i\right]\right)^{2}
\]

\begin{figure}
\begin{centering}
\includegraphics[width=0.7\columnwidth]{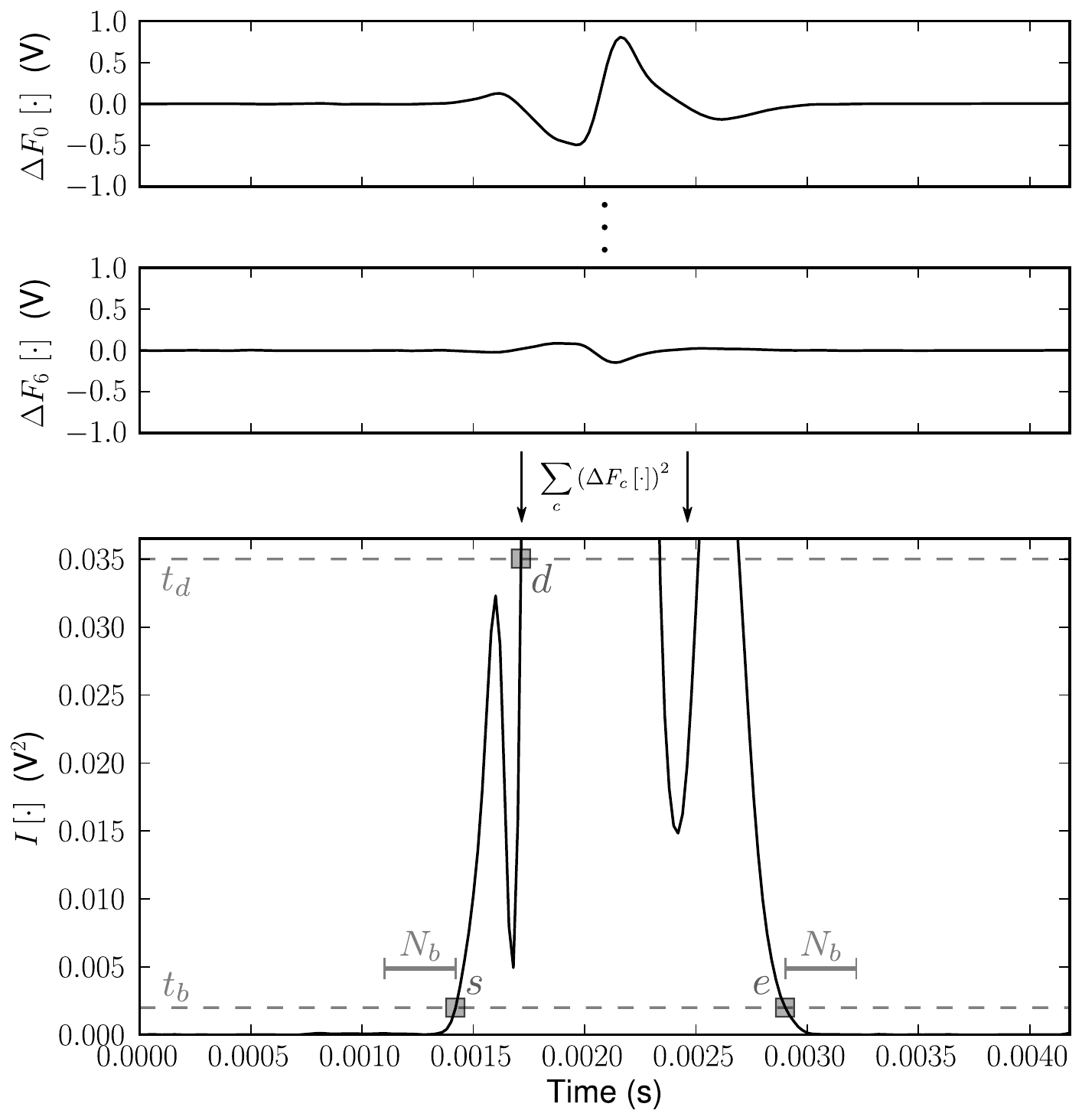}
\par\end{centering}

\caption{\label{fig:Method_Segments}Signals $\Delta F_{c}\left[\cdot\right]$,
obtained by filtering and differentiating experimental data collected
from every channel $c$, are squared and summed to compute $I\left[\cdot\right]$.
When $I\left[\cdot\right]$ surpasses a $t_{d}$ threshold, a segment
of EOD activity is detected. To establish the $s$ (and $e$) segment
boundaries, we look for a $N_{b}$ idle time occurring before (after)
the signal goes below a $t_{b}$ threshold.}
\end{figure}

Segment boundaries (starting and ending time instants $s$ and $e$)
are determined by iterating over samples of the summed signal $I\left[\cdot\right]$,
in both directions, starting at the position $I\left[d\right]$ where
detection ($I\left[d\right]>t_{d}$) took place, until a certain number
$N_{b}$ of contiguous samples ahead of the boundary is found to be
below a minimum value $t_{b}$, which is set to be lower than the
detection threshold (such that $t_{b}<t_{d}$). This procedure is
shown in Figure \ref{fig:Method_Segments} and described by the equations
below.

\[
\max\left\{ \begin{aligned}s\in\mathbb{N},\, s<d\::\: & \forall i\in\mathbb{N},\,1\le i\le N_{b}\\
\, & \rightarrow I\left[s-i\right]<t_{b}
\end{aligned}
\right\} 
\]

\[
\min\left\{ \begin{aligned}e\in\mathbb{N},\, e>d\::\: & \forall i\in\mathbb{N},\,1\le i\le N_{b}\\
\, & \rightarrow I\left[e+i\right]<t_{b}
\end{aligned}
\right\} 
\]

The parameter $t_{d}$ should be set just below the peak caused on
$I\left[\cdot\right]$ by an EOD emitted when fish is at the position
which leads to the minimum amplitude of acquired signals, i.e. at
the middle point of the aquarium; $t_{b}$ should be adjusted above
the maximum noise floor; and $N_{b}$ should be greater than the number
of contiguous $I\left[\cdot\right]$ samples of an EOD that might
be below $t_{b}$. We have developed a graphical user interface for
adjusting these parameters which shows an interactive graphic similar
to the one presented in the figure. For our experimental setup, we
have chosen $t_{d}=0.06\;\mathrm{V}^{2}$, $t_{b}=0.0012\;\mathrm{V}^{2}$
and $N_{b}=16$.

The filtering operation carried before segmentation is meant to reduce
the susceptibility to noise of $s$ and $e$ measurements. Filter
choice is not critical as long as the spiking shape of the EOD is
preserved. The purpose of the differentiation is to remove any DC
component that might be left on the signal due to offsets in the acquisition
system, avoiding spurious detections. Both operations are carried
solely to aid EOD segmentation and their results are not used by the
next steps of the algorithm. A less redundant approach is to replace these 
two operations by a signal reconstruction from the second leaf of the third
level of the wavelet transform employed in Subsection
\ref{sub:Feature-extraction} (equivalent to a sequence of low-pass, low-pass again, and finally high-pass filtering operations). This produces a frequency response which is
very close to that of the already described $\Delta F_{c}\left[\cdot\right]$
signal, with the advantage of allowing the computed wavelet signal
components to be subsequently reused.

In order to compute feature vectors from a segment containing a single
EOD, a signal window of a fixed length is needed. We choose a length
$l_{w}$ large enough to accommodate any single EOD, and center the
segments into windows by computing window boundaries $s'=\nicefrac{\left(s+e-l_{w}\right)}{2}$
and $e'=s'+l_{w}$ given the segment boundaries $s$ and $e$. In
\textit{Gymnotus} sp., pulse duration typically lies in the range
of 1.8 to 2.2 ms, thus we choose $l_{w}=128$ (2.56 ms) when acquiring
data at 50 kHz.

\subsubsection{Feature extraction\label{sub:Feature-extraction}}

One set of features is independently computed for each channel $c$
of the original signal $A_{c}\left[\cdot\right]$, as it was before
filtering and differentiation. We have selected feature extraction
schemes which produce output less sensitive to fish position than
the time-domain signals. A simple approach is to compute the Fourier
transform $\tilde{A_{c}}\left[\cdot\right]=\mathfrak{F}\left\{ A_{c}\left[s'\cdots e'\right]\right\} $
of pulse signal windows, take the complex magnitude $\bigl|\tilde{A_{c}}\left[\cdot\right]\bigr|$
of the resulting values --- which represents the amplitude of each
frequency component without phase information --- and finally normalize
the vector by dividing all the values by the largest one ($\bigl|\tilde{A_{c}}\left[\cdot\right]\bigr|/\max_{i}\bigl|\tilde{A_{c}}\left[i\right]\bigr|$).

Nonetheless, it has been reported \citep{paintner2003electrosensory}
that some species of fish are able to distinguish between pairs of
different artificial EOD pulses possessing the same amplitudes of
the frequency spectrum. In the cited work, the artificial EODs were
constructed by a superposition of two time-shifted components. Therefore,
the EODs differed only in the phase of the frequency spectrum and
would be indistinguishable by the aforementioned means. We hypothesized
that if fish can sense these variations in artificial EODs, real discharges
could conceivably contain somewhat time-localized features, just like
the time-shifted components of the artificial signal. For example,
some range of frequencies could be particularly more prevalent at
the beginning of the EODs emitted by a specific individual. This kind
of information is useful for the classifier and can be extracted employing
time-frequency techniques such as wavelet transforms.

Wavelet packet transforms build decomposition trees which, at each
level, refine signal localization in frequency while loosing localization
in time. We have chosen the dual-tree complex wavelet packet transform
(DT-CWPT) because it is nearly shift-invariant, although it can sense
time shifting when the decomposition level allows for sufficient time
localization. The ordinary discrete wavelet packet transform (DWPT)
does not hold this property and is susceptible to artifacts induced
by phase variation at all levels of decomposition. This is an undesirable
effect, since as far as possible we only want to take into account
phase differences in EOD signals due to factors other than fish position.

We have implemented DT-CWPT using 20-tap Q-shift filters \citep{qshiftdesign}
and Dau\-be\-chies wavelet filters with seven vanishing moments. The dual
trees are combined to obtain a single tree comprising the complex
magnitude of each component. As the transform is computed on windows
of $l_{w}=128$ samples, trees decompose the signals into $\log_{2}\bigl(l_{w}\bigl)=7$
levels, each one containing 128 components. We independently normalize
each level, by dividing all values by the largest one present in
the same level of the tree.

In short, the DT-CWPT and Fourier transforms provide, respectively,
$l_{w} \cdot \allowbreak \log_{2}\bigl(l_{w}\bigl) \allowbreak = \allowbreak 896$
and $\nicefrac{l_{w}}{2} \allowbreak = \allowbreak 64$
signal components which, after normalized, can form the feature vectors.

\begin{figure}
\begin{centering}
\includegraphics[width=1\columnwidth]{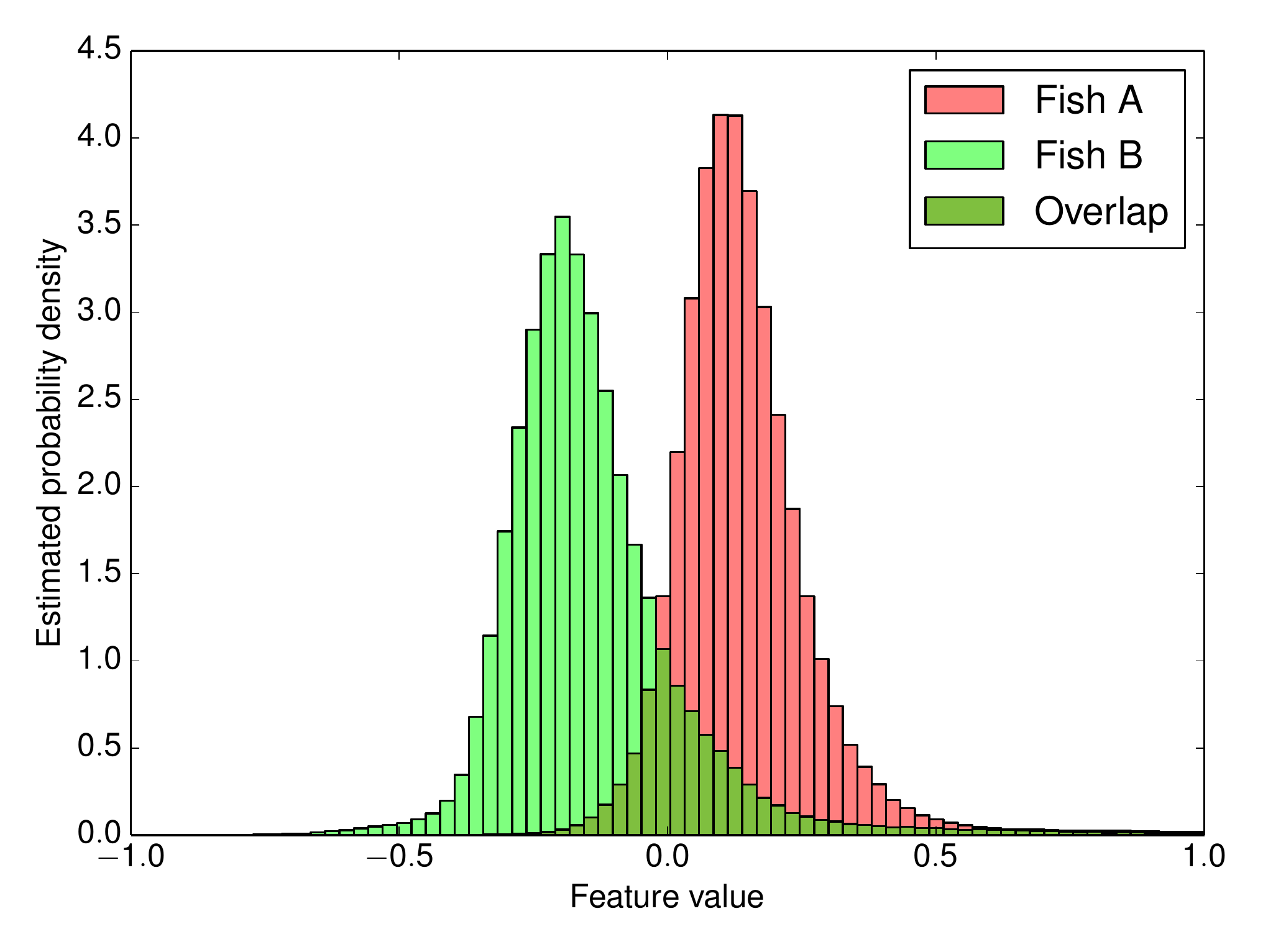}
\par\end{centering}

\caption{\label{fig:Method_Histogram_Overlap}The histogram of a Fourier or
wavelet component can be employed to estimate the probability distribution
of its values. The overlapping area between histograms of the component
in two distinct fish is used as a measure of the uncertainty of classifying
EODs if only this single feature was known. We adopt the straightforward
feature selection approach of choosing the components presenting the
least histogram overlapping area.}
\end{figure}

\subsubsection{Feature selection\label{sub:Feature-selection}}

The number of signal components provided by DT-CWPT and Fourier transforms
is large (960 in total). If all these components were used as features,
training times would be excessively long and the model would likely overfit.
Therefore, we select only 20 signal components to constitute the feature
vectors that feed the classifier.

To aid feature selection, we build two histograms for each signal component,
one corresponding to each fish, as illustrated in
Figure \ref{fig:Method_Histogram_Overlap}.
These histograms have the same bin intervals, so that the overlapping
area, highlighted in the figure, can easily be calculated by integrating
over the least of two bin heights at each point. The minimum overlapping
area (zero) would be obtained in the ideal situation where training
set EODs could be perfectly classified using only this single feature.
The maximum area (one) would be reached in the worst case, when probability
distributions of the signal component are almost the same in both
fish, meaning the component is worthless for classification if used
alone.

Although signal components cannot be considered independent from each
other, we assume for simplicity that selecting the ones presenting
the least histogram overlapping area makes a good feature set. For
the problem at hand, we verified empirically that this simple form
of filter approach \citep{John94irrelevantfeatures} for feature
selection works well.

\subsubsection{Supervised classifier\label{sub:Supervised-classifier}}

SVM classifiers were adopted because they frequently produce good
results for a variety of problems \citep{Meyer2003169}. Furthermore,
mature and optimized SVM software libraries are readily available
\citep{Chang:2011:LLS:1961189.1961199}. Each feature supplied to
the classifier is rescaled to the $\left[-1,1\right]$ interval, in
order to prevent dominance of the features spanning the larger numeric
ranges. Our implementation uses by default the gaussian radial basis
function (RBF) kernel, usually considered a good first choice, since
depending on its parameters it can also behave like linear or sigmoid
kernels \citep{hsuSVMGuide}.

Well-formed EOD signals (i.e. those which are non-saturated and have
a good SNR, as verified by checking the EOD amplitude) obtained from
a single fish during training stage are randomly distributed into
three sets, each one containing on the order of $10^{4}$ EODs if
training stage was carried for about 15 minutes per fish. One of them,
the training set, is used to train the SVM model; another, the validation
set, to count the number of errors throughout a grid search intended
to find optimal RBF kernel and soft-margin parameters ($\gamma$ and
$C$); and the third, the testing set, to estimate final SVM performance
on single fish discharge classification.

Once the SVM model is trained, data collected from the main experiment
can be processed. Signal segments obtained from this data may contain
either an EOD emitted by a single fish or EODs fired by both fish
almost at the same time. Diverse criteria are checked to establish
with a high true negative rate (specificity) whether a certain segment
contains a single EOD.

First, we dismiss segments whose length ($e-s$) is greater than $\max\{\bar{l_{1}}+\sigma_{l_{1}},\bar{l_{2}}+\sigma_{l_{2}}\}$,
where $\bar{l_{j}}$ is the mean and $\sigma_{l_{j}}$ is the standard
deviation of the length of signal segments present in training data
collected from the $j$-th fish alone. We also reject segments where
less than $N_{r}$ of the available channels captured well-formed
(good SNR, non-saturated) signals, and only consider pairs of adjacent
segments such that the time interval between their starting instants
is less than the minimum discharge period attainable by a single fish,
in which case the EODs were probably emitted by different individuals.

Finally, the SVM model is employed to compute Platt \citep{Platt99probabilisticoutputs,PlattProbLibSVM}
probability estimators $p_{c}$ for every channel $c$ containing
a well-formed signal during each segment. If one of the products $\prod_{c}p_{c}$
or $\prod_{c}(1-p_{c})$ is above a certain threshold $t_{p}$, the
associated segment is marked as being emitted by the first or by the
second fish, respectively. Segments pertaining to an adjacent pair
must have marks corresponding to different fish, otherwise the classification
is deemed as incorrect.

We typically adopt $N_{r}=2$ and $t_{p}=0.95$. The larger the values
of $N_{r}$ and $t_{p}$, the greater the overall specificity of the
single fish classifier. We do not recommend choosing $N_{r}>2$ because
it is rare to observe a well-formed EOD in more than two channels
at the same time. Choosing $N_{r}=1$ can be effective if one fish
stayed for several seconds of the experiment in a position of the
aquarium where a well-formed EOD is captured by a single channel,
but in these cases we recommend to compensate specificity by increasing
$t_{p}$ to values in the order of $0.99$. Classification errors
on this step of the algorithm will propagate to the next pass (Subsection
\ref{sub:Continuity-constraint}), therefore forcing a high specificity
is important to get a final discrimination error in the order of parts
per million even though the classification error of a single channel
can reach hundreds of parts per million (see Table \ref{tab:Results_10fold}).

\subsubsection{Continuity constraint\label{sub:Continuity-constraint}}

\begin{figure*}
\begin{centering}
\includegraphics[width=1\textwidth]{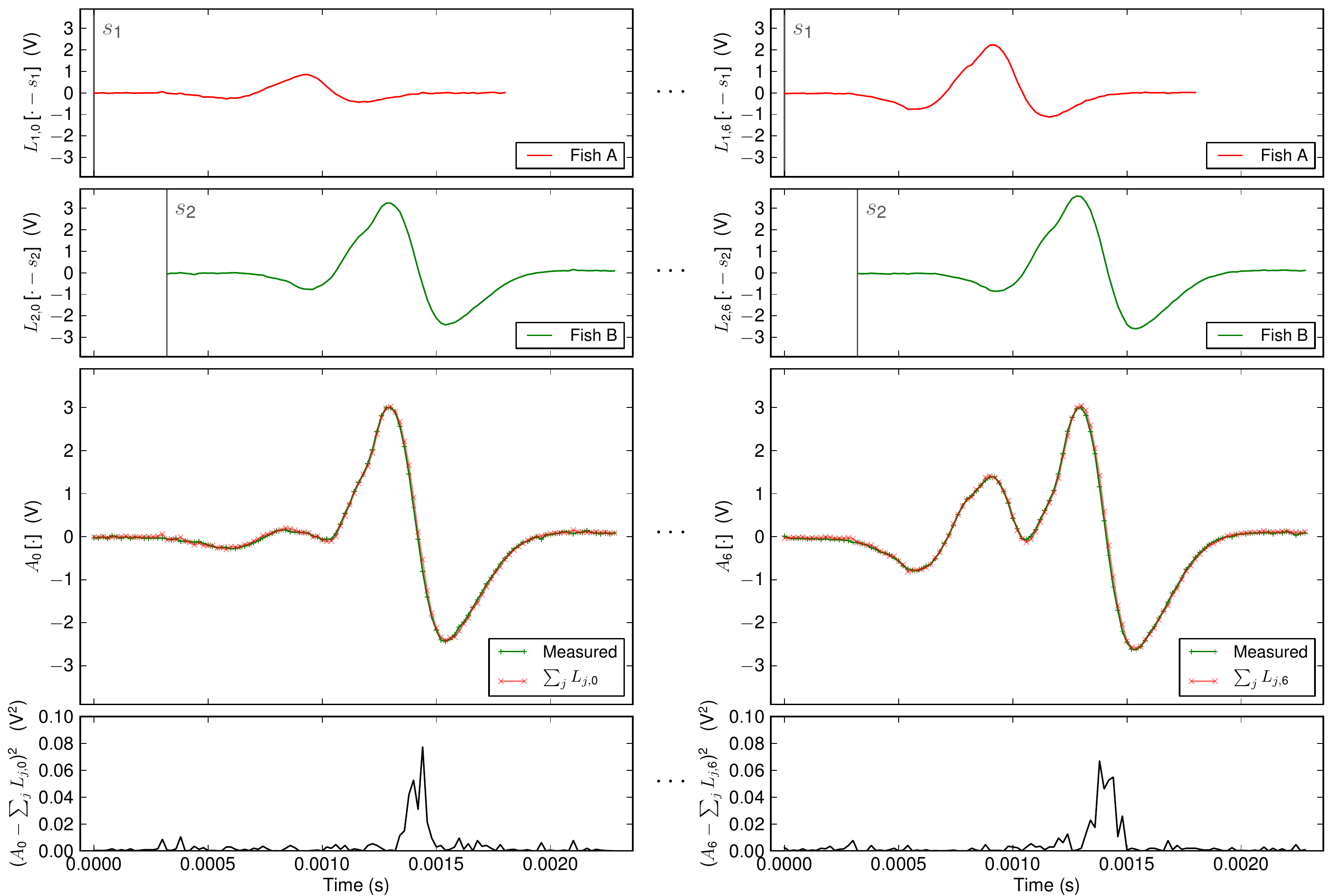}
\par\end{centering}

\caption{\label{fig:Method_Argmin} A segment of the $A_{c}\left[\cdot\right]$
signals may contain EODs emitted by both fish almost at the same time.
In order to identify the positions $s_{j}$ where EODs emitted by
the $j$-th fish start within the segment, we exploit the continuity
property of EOD waveforms emitted by each fish. The figure is organized
as a grid, where each column comprises graphics presenting data related
to a certain channel $c$ of the signal. The first two lines show
the $L_{j,c}$ vectors, which contain the last EOD previously recognized
as being emitted by the $j$-th fish, displaced to the positions $s_{j}$
being evaluated. In the third line, both displaced vectors are summed,
and the resulting $\sum_{j}L_{j,c}$ are compared to the measured
$A_{c}$ signals. In this example, no saturation occurs, thus there
is no need to apply the $\Xi$ function defined in the text. The instants
$s_{j}$ are chosen to minimize the sum of all squared distances displayed
in the fourth line of the figure.}
\end{figure*}

Whenever an adjacent pair of segments fulfilling the aforementioned
criteria is detected, for both $j\in\left\{ 1,2\right\} $, vectors
$L_{j,c}\left[\cdot\right]$ are initialized with signals from every
channel $c$, well-formed or not, belonging to the segment classified
as containing a single EOD emitted by the $j$-th fish. Then, a continuity
constraint is imposed to allow recognizing EODs present in subsequent
signal segments, until the next criteria-satisfying adjacent pair
is found. We assume that EOD waveforms collected from an individual
vary smoothly, because fish position in the aquarium is a continuous
function of time. This assumption is similar to an established approach
\citep{Snider1998155} in spike sorting literature for solving the
electrode drift problem.

Dissociation of signal segments containing EODs from both fish is
thus formulated as an Euclidean distance minimization problem. Starting
time instants $s_{j}$ of EODs emitted by the $j$-th fish are found
by shifting the $L_{j,c}\left[\cdot\right]$ vectors to the positions
$s_{j}$ being tested, summing shifted vectors corresponding to different
fish, and comparing the resulting signal to the current segment $A_{c}\left[s\cdots e\right]$,
as defined by the following equation and illustrated in Figure \ref{fig:Method_Argmin}.

\[
\argmin_{s_{1},\, s_{2}}\sum_{i,\, c}\left(A_{c}\left[i\right]-\Xi\left(L_{1,c}\left[i-s_{1}\right]+L_{2,c}\left[i-s_{2}\right]\right)\right)^{2}
\]

In our notation, $L_{j,c}\left[i\right]$ evaluates to zero whenever
$i$ is outside of the bounds ($0\le i<l_{w}$). The function $\Xi\left(V\left[\cdot\right]\right)$
roughly models the effect of differential amplifier output signal
saturation. If the maximum and minimum output voltage swing values
are given by $v_{sh}$ and $v_{sl}$, then $\Xi$ can be defined as
follows.

\[
\Xi\left(V\left[k\right]\right)=\begin{cases}
v_{sh}\text{,}\: & \text{if }\, V\left[k\right]\ge v_{sh}\\
v_{sl}\text{,}\: & \text{if }\, V\left[k\right]\le v_{sl}\\
V\left[k\right]\text{,}\: & \text{otherwise}
\end{cases}
\]

When testing different $s_{j}$ values, we allow $s_{1}$ to take
a single value where $L_{1,c}\left[i-s_{1}\right]$ always evaluates
to zero, meaning $L_{2,c}\left[\cdot\right]$ alone is compared to
$A_{c}\left[s\cdots e\right]$, and vice versa. This way, the Euclidean
distance minimization step can also be employed to detect signal segments
containing a single EOD which were dismissed by the previous criteria
due to its high specificity. When these segments are identified, the
algorithm updates the $L_{j,c}\left[\cdot\right]$ vectors corresponding
to the recognized fish $j$, in order to reflect the latest waveforms.

\section{Results\label{sec:Results}}

\begin{table}
\begin{centering}
\begin{tabular}{|c|c|}
\hline 
Dyad & Cross-validation accuracy\tabularnewline
\hline 
\hline 
1 & 99.9828\%\tabularnewline
\hline 
2 & 99.9998\%\tabularnewline
\hline 
3 & 99.9366\%\tabularnewline
\hline 
4 & 99.9993\%\tabularnewline
\hline 
5 & 99.9064\%\tabularnewline
\hline 
6 & 99.9993\%\tabularnewline
\hline 
\end{tabular}
\par\end{centering}

\caption{\label{tab:Results_10fold}Results of a 10-fold cross-validation test
carried applying our SVM based method on the entire data set collected
during the training stage, for six different dyads.}
\end{table}

\begin{figure*}
\begin{centering}
\includegraphics[width=1\textwidth]{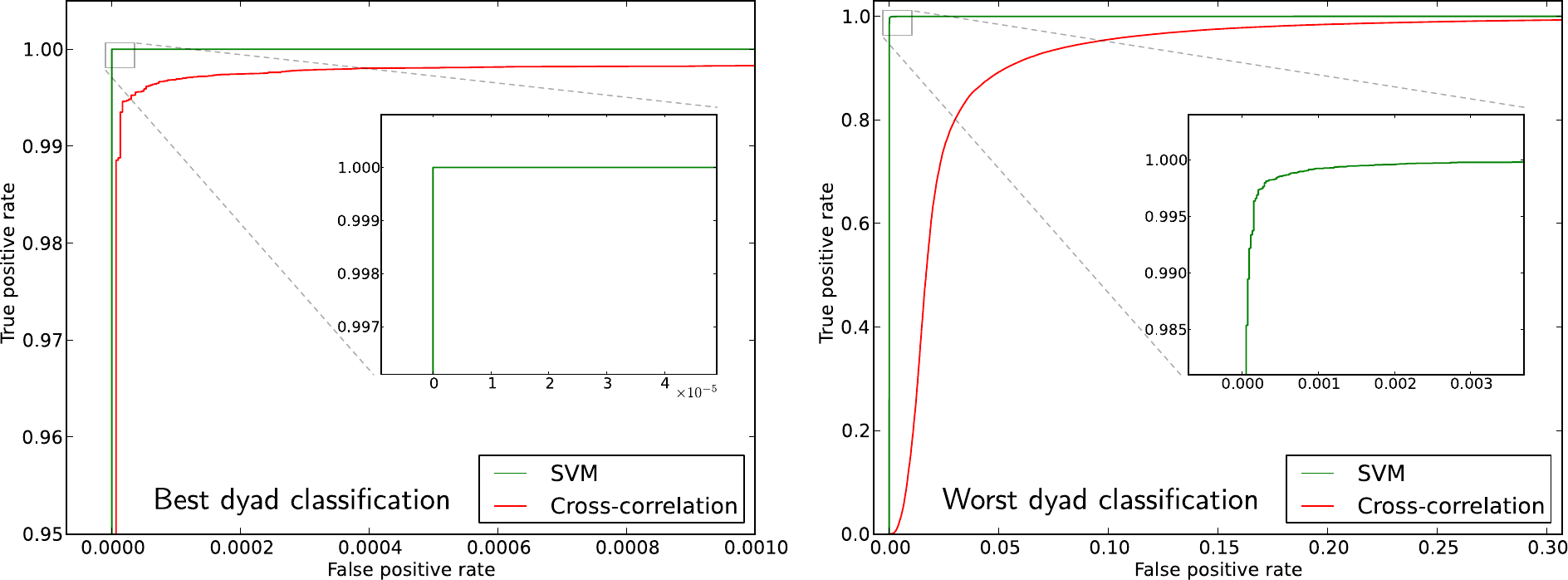}
\par\end{centering}

\caption{\label{fig:Results_ROC}Receiver operating characteristic (ROC) curves
comparing SVM and cross-correlation methods for classifying signal
windows of the testing set. Each window contained a single EOD emitted
by an individual of a dyad (A and B fish). True positive rate denotes
the ratio of EODs emitted by fish A which were correctly identified.
Likewise, false positive rate corresponds to the ratio of incorrectly
identified EODs emitted by fish B. Graphics were plotted for two different
dyads, the ones which gave best and worst classification accuracy,
respectively, among the six dyads used during experiments.}
\end{figure*}

First, we evaluated the SVM model alone regarding its ability to classify
signal windows containing a single EOD. We present in Table \ref{tab:Results_10fold}
the results of a standard 10-fold cross-validation test conducted
over all of the data collected during the training stage. In other
words, not taking into account the division into training, validation
and testing sets mentioned in Subsection \ref{sub:Supervised-classifier}.

We also compared our SVM approach to the cross-correlation method
described in Subsection \ref{sub:Cross-corr-method}, by plotting
the receiver operating characteristic (ROC) curves shown in Figure
\ref{fig:Results_ROC}. The data set used to obtain these curves is
disjoint from the ones adopted to train the classifiers, i.e. curves
were constructed based on classification results of the testing set,
employing models trained only with the training set and with hyper-parameters
optimized using the validation set. We plotted curves both for the
best and for the worst case, corresponding to the dyads numbered 2
and 5 in Table \ref{tab:Results_10fold}, respectively. ROC results
show that our SVM approach performs consistently better than the cross-correlation
method. In the worst case data set, where SVM displayed particularly
superior accuracy compared to cross-correlation, EODs emitted by different
individuals had almost the same pulse duration, a situation on which
the cross-correlation method is known \citep{ISI:000248304900011}
to work poorly.

Next, we analyzed data collected during the validation procedure (Subsection
\ref{sub:Validation-procedure}). By comparing direct electrode measurements
with the discrimination algorithm results, we found a single non-detected
EOD among $7.4\times10^{5}$ pulses, resulting in an error rate in
the order of two parts per million EODs. For the discharge rate of
a typical \textit{Gymnotus} sp., this represents a mean interval of
one to two hours of data collection between errors.

\begin{figure*}
\begin{centering}
\includegraphics[width=1\textwidth]{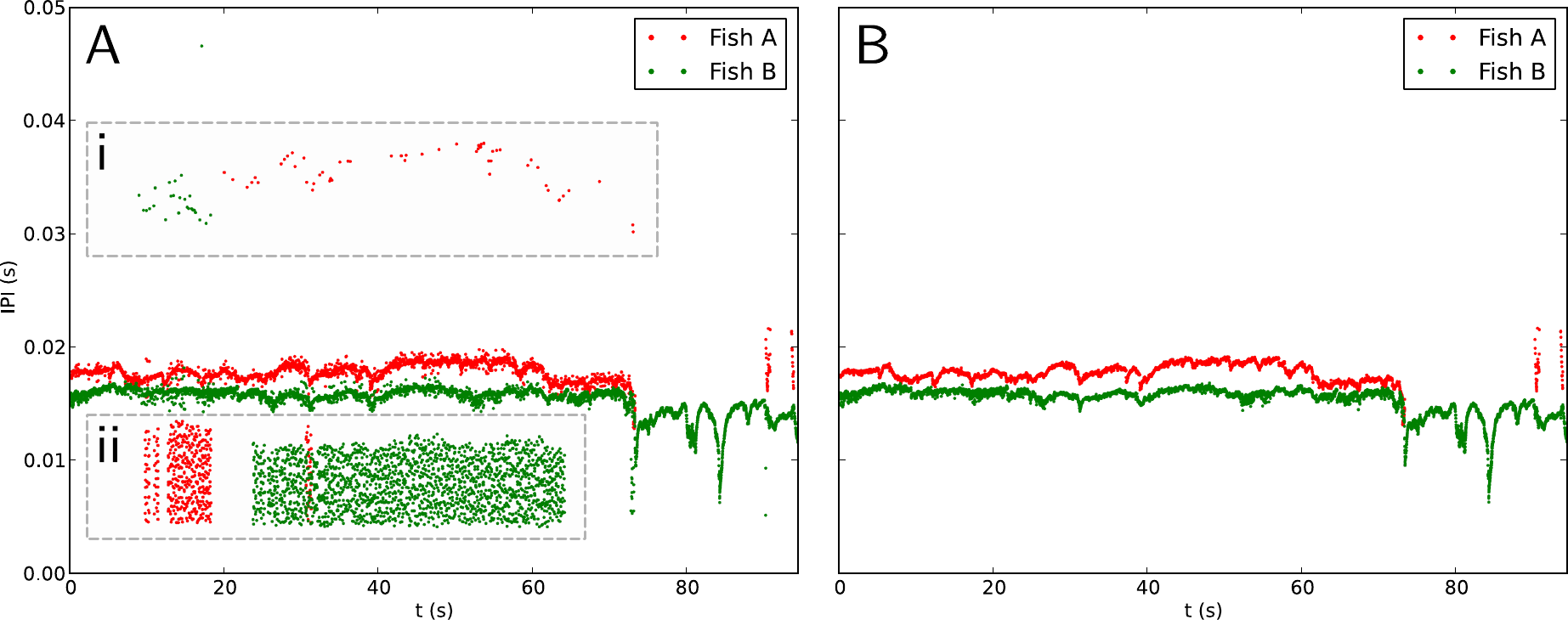}
\par\end{centering}

\caption{\label{fig:Results_IPI}Inter-pulse interval (IPI) graphics plotted
using two different discrimination algorithms, given the same data
collected from a freely swimming fish dyad. On correct discrimination,
IPI is expected to be a piecewise continuous function. Signals from
which we computed IPI were not used during training. Sub-figure (A)
shows results from a previous version of our algorithm which used
the SVM alone (without applying the waveform continuity constraint),
and therefore had difficulties when both fish emitted EODs almost
at the same time or when some channel saturated. Discrimination errors
can be easily pinpointed: region (i) contains missed EODs, and is
located approximately at the double IPI of the baseline; region (ii)
contains false positives, presenting IPIs below the baseline. Sub-figure
(B) displays results from our fully implemented algorithm, as described
in this paper. No errors can be pinpointed in its IPI graphic.}
\end{figure*}

Finally, we observed inter-pulse interval (IPI) graphics plotted using
results of the discrimination algorithm when applied to data collected
with both fish of a dyad freely swimming, absent of any directly attached
electrodes. As individuals were not restrained in any way during this
test, it was conducted the closer way possible to a natural setting.
Even though direct measurements are not available in this sort of
experiment, a kind of validation can still be carried, because the
IPI curve for a fish is expected to be piecewise continuous. An individual
can stop emitting EODs for a short period of time, but when it is
emitting EODs, the pulse rate (and thus the IPI) varies smoothly.

The left panel of Figure \ref{fig:Results_IPI} shows the IPI obtained
employing a previous version of our algorithm, which did not incorporate
the waveform continuity constraint discussed in Subsection \ref{sub:Continuity-constraint}.
It simply classified signal windows using SVM and treated a window
as containing pulses from both fish whenever signals of distinct channels
were recognized as emitted by different fish. This led to a numerous
amount of false positives when signal saturation occurred, and to
some missed EODs when both fish fired almost at the same time. We
show these results solely to illustrate that discrimination errors
can be easily spotted in an IPI plot. One missed EOD appears as a
point located at the double IPI of the baseline, as portrayed in the
region (i) of the figure. Similarly, any EODs detected by the algorithm
which did not really exist appear below the baseline, as displayed
in region (ii).

On the other hand, the right panel of the figure shows the results
of the fully implemented method, as described in Subsection \ref{sub:Proposed-algorithm}.
No errors can be pinpointed in this plot, as the IPI varies piecewise
continuously for each fish.

However, we stress the fact that this IPI continuity property is not
exploited by the discrimination algorithm. The continuity constraint
used by our algorithm pertains only to the EOD waveform, which is
a function of the spatial location of the fish, and is in no way related
to the pulse rate. Therefore, no cyclic argument exists, and the fact
that we observe smooth IPI curves when plotting the data obtained
with our algorithm is one more reason to trust its correctness.

\section{Conclusions\label{sec:Conclusions}}

We have presented a method able to accurately recognize the individual
which emitted each electric organ discharge (EOD) during experiments
conducted with freely swimming \textit{Gymnotus} sp. dyads. The obtained
data is useful for analyzing and studying communication protocols
employed by the animals in a range of interesting situations, like
mating, dominance relation establishment and territorial dispute,
besides being important for shedding more light on fundamental issues,
such as efficiency and redundancy of communication signal coding,
jamming avoidance response \citep{CapurroJAR} and communication channel
multiplexing mechanisms which might be present in these animals.

Our method requires only a simple experimental setup, consisting of
an array of fixed electrodes, conventional operational amplifier circuits
for conditioning signals, and a data acquisition system. Electrodes
can be affixed to an aquarium or be mounted onto a structure which
can be installed inside a fishpond. Unlike procedures carried in previous
studies, no cameras are needed, allowing experiments to be easily
carried in turbid waters, which are a common habitat of these animals
\citep{Baffa1992591}. Also, we are able to reliably process a large
amount of collected data, which is essential for the attainment of
more faithful results when applying information theoretic and statistical
approaches to analyze communication signals.

Experiments carried out with \textit{Gymnotus} sp. gave outstanding
discrimination results, therefore we believe this method could be
applied to other species of pulse-type electric fish, although it
remains to be attested if those species present individual distinguishable
signatures which could be identified by our support vector machine
(SVM) based classifier. Additionally, whilst we have devised and implemented
the method only for dealing with a fish dyad, it can be naturally
extended to process signals collected from more than two individuals.
Notwithstanding, naively expanding the terms of the proposed continuity
constraint for handling more individuals would be too much computationally
expensive due to the processing power needed to deal with the large
amounts of data. Therefore, some heuristics would need to be developed
in order to reduce the optimization search space.

As a future work, we plan on implementing this algorithm in field-pro\-gram\-ma\-ble
gate array (FPGA) hardware, employing a shift-reg\-is\-ter-like architecture
for the continuity constraint step. Such a device would allow us to
obtain discriminated fish EOD instants with low latency and jitter
characteristics. This would be suitable for conducting a new range
of re\-al-time closed-loop experiments such as involving a third artificial
fish communicating with the dyad.

\section*{Acknowledgements}

This work was supported by a grant from the\foreignlanguage{brazil}{
\hyphenation{CAPES}\hyphenation{CNPq}\hyphenation{FAPESP}CAPES --
Coordenação de Aperfeiçoamento de Pessoal de Nível Superior} -- Brazilian
agency. We also acknowledge the \foreignlanguage{brazil}{FAPESP --
Fundação de Amparo à Pesquisa do Estado de São Paulo} -- and \foreignlanguage{brazil}{CNPq
-- Conselho Nacional de Desenvolvimento Científico e Tecnológico}
-- agencies for their financial support for past and future projects
related to this work. We thank Lirio O. B. Almeida, Roland Köberle,
Rafael T. Guariento and Krissia Zawadzki for reviewing our original
manuscripts.

\bibliographystyle{model1-num-names}
\phantomsection\addcontentsline{toc}{section}{\refname}\bibliography{paper}

\end{document}